\documentclass[preprint,12pt
,flushrt]{elsarticle}
\usepackage{graphicx}% Include figure files
\usepackage{dcolumn}% Align table columns on decimal point
\usepackage{bm}% bold math
\usepackage{latexsym}
\usepackage{color}
\usepackage{natbib}
\usepackage{amssymb}
\usepackage{aas_macros}
\usepackage{amsmath,amsfonts,latexsym,amstext,amsbsy,amsthm}
\journal{Physics Letters B}

\begin{document}
\begin{frontmatter}

\title{Origin of Intense Magnetic Fields Near Black Holes Due to Non-Minimal Gravitational-Electromagnetic Coupling.}% Force line breaks with \\

\author{Rafael S. de Souza\corref{cor1}}
\ead{Rafael@astro.iag.usp.br}
\author{Reuven Opher\corref{cor2}}%
 \ead{Opher@astro.iag.usp.br}
\cortext[cor1]{IAG, Universidade de S\~{a}o Paulo, Rua do Mat\~{a}o 1226, Cidade
Universit\'{a}ria, CEP 05508-900, S\~{a}o Paulo, SP, Brazil.}
\cortext[cor2]{IAG, Universidade de S\~{a}o Paulo, Rua do Mat\~{a}o 1226, Cidade
Universit\'{a}ria, CEP 05508-900, S\~{a}o Paulo, SP, Brazil.}

\begin{abstract}
The origin of magnetic fields in astrophysical objects is a
challenging problem in astrophysics. Throughout the years, many
scientists have suggested that non-minimal
gravitational-electromagnetic coupling (NMGEC) could be the origin
of the ubiquitous astrophysical magnetic fields.  We investigate
 the possible origin of intense magnetic fields  by
  NMGEC near rotating  black holes,  connected with  quasars   and gamma-ray bursts.
   Whereas these intense magnetic fields
   are difficult to explain astrophysically, we find that they are easily explained by NMGEC.

\end{abstract}


\end{frontmatter}

\section{Introduction}

Cosmic magnetic fields pervade the Universe. However, their origin
is one of the most challenging problems in modern astrophysics
\citep[e.g.,][]{ree87,kro94,raf2011,raf10b,raf10c,rod10,lag10,raf08}. Various authors have suggested a
gravitational origin of the magnetic fields in rotating celestial
bodies.  In particular, a number of studies have been made on
nonminimal gravitational electromagnetic coupling (NMGEC). It has
been motivated, in part, by the Schuster-Blackett (S-B) conjecture,
which suggests that the magnetic fields in planets and stars arise
due to their rotation  \citep{schu11}. In this scenario, neutral mass
currents generate magnetic fields, implying the existence of a
non-minimal coupling between gravitational and electromagnetic
fields. An early attempt to encompass the S-B conjecture in a
gravitational theory was made  by  Pauli in the 1930s \citep{pau33}.
During the 1940s and 50s, after  Blackett  resuscitated the
conjecture \citep{bla47}, many authors,  such as \citet{ben49},
%\citet{pap50}%,
 and \citet{luc51},  also attempted to encompass it  in a
gravitational theory. 
Later, in the eighties, \citeauthor{bar85}
also studied the NMGEC conjecture \citep{bar85}.
 The majority of these studies were based on the five-dimensional
Kaluza-Klein formalism. This formalism was used in order  to
describe a unified theory of gravitation and electromagnetism with
NMGEC in such a way that the S-B conjecture is obtained.
\citet{oew97} proposed that the $B \sim 10^{-6}-10^{-5}$ G magnetic
field in spiral galaxies is directly obtained from NMGEC. 
\citet{mik95} showed that Mikhail and Wanas tetrad theory of gravitation  (MW) \citep{mik77,mik81} predicts the S-B conjecture of NMGEC.
More recently \citep{rafael10} have used the NMGEC-MW formalism to explain the magnetic fields in the central engine of gamma-ray bursts. 
  In this paper, we investigate  the possibility that NMGEC is the origin
of the intense magnetic fields near rotating  neutron stars and
black holes, connected with  quasars, and gamma ray
bursts.

\section{Basic Features of the Model}

NMGEC suggests the following relation between the angular momentum
\textbf{L} and the magnetic dipole moment \textbf{m}:
\begin{equation}
\textbf{m}=\left[\beta \frac{\sqrt{G}}{2c}\right]\textbf{L},
\end{equation}
where $\beta$ is a constant, G  the Newtonian constant of
gravitation, and c is the speed of light.  The angular momentum
\textbf{L} is

\begin{equation}
\mathbf{ L}= I \mathbf{ \Omega},
\end {equation}
where $\mathbf{ \Omega} = 2 \pi P^{-1}$ is the angular velocity, P
the rotational period, and I is the moment of inertia. The
 dipole moment $\vec{\mathbf{m}}$is related to the magnetic field $\vec{\mathbf{B}}$ by
\begin{equation}
\vec{\mathbf{B}} = \frac{3(\vec{\mathbf{m}}\cdot
\vec{\mathbf{r}})\vec{\mathbf{r}}-\vec{\mathbf{m}}\vec{\mathbf{|r|}}^{2}}{\vec{\mathbf{|r|}}^{5}},
\end{equation}
where $\vec{\mathbf{r}}$ is the distance from $\vec{\mathbf{m}}$ to the point at which $\vec{\mathbf{B}}$ is measured.

\section{Quasars}

Supermassive black holes are generally believed to be the power
sources of quasars and other active galactic nuclei.  Apart from its
mass, the other fundamental properties of an astrophysical black
hole
 are its charge and, in particular, its spin. A spinning Kerr black hole has a greater radiative
 efficiency than that of a non-rotating Schwarzschild black hole. Both are expected to have
 negligible charge due to the high conductivity of the surrounding plasma. \citet{wan06} estimated the average
 radiative efficiency of a large sample of quasars, selected from
 the Sloan Digital Sky Survey, by combining their luminosity
 and their black hole mass functions.  They found that quasars have an average radiative
 efficiency of $\sim 30 \% - 35\%$ over the redshift interval $0.4
 < z < 2.1$. This strongly suggests that the Kerr
 black holes are rotating very rapidly with approximately maximum  angular
 momentum, which remains roughly constant
 with redshift. The inferred large spins and their lack of
 significant evolution with redshift are in agreement with the predictions of
  semianalytical models of hierarchical galaxy formation (\citep{vol05}, \citep{sol82}, \citep{wan06}).
   In these  models,
 black holes gain most of their mass through accretion.

   Using the  rotation measures (RMs) of high redshift galaxies, \citet{pen00}
obtained an estimate of the accretion disk magnetic field in the
region where polarized optical radiation is generated.
 Assuming that
the magnetic flux is conserved and that the optical radiation is
emitted from  the accretion disk  in the region $\sim 10^{3} r_{g}$,
(where $r_{g}$ is the gravitational radius of a supermassive black
hole), we obtain the following estimate of the accretion disk
magnetic field in the generation region of the optical radiation:

\begin{equation}
B \sim 2 \times 10^{3}  \ (RM/10^{3})(10^{8}M_{\odot}/M_{BH})^{2} G
\end{equation}
\citep{pen00}. The field strength given by (6) for quasar accretion
disks was found to be $\sim 150-300$ G \citep{pen00}.

We can compare this value with the NMGEC prediction for magnetic
fields in quasars.  Using (1) and taking
 $M_{BH} \sim 10^8 M_{\odot}$ and  $\beta \sim 1$,
we obtain B $\sim 10^{9}$ G near $r_{g}$.
 Assuming that the magnetic flux produced by NMGEC is conserved as
 it expands from $r_{g}$ to $10^{3} r_{g}$ (decreasing as $1/r^{2}$), we
 obtain $B \sim 10^{3}$ G at $r \sim 10^{3} r_{g}$, which is in good agreement with
 the quasar accretion disk  magnetic field obtained from (6).

\section{Gamma-Ray Bursts}

Magnetic fields are very important in   Gamma-Ray Bursts
(GRBs)\citep{pir04}. It is generally accepted that  the observed
afterglow is
 produced by synchrotron emission which involves magnetic fields.  Synchrotron radiation
 is also the  best model for
 prompt $\gamma$-ray emission.
 The relativistic outflow is a Poynting flux (with negligible baryon
 content) \citep{pir04}. A natural way to produce the Poynting flux is
 by magnetic reconnection.

  The magnetic field required  for the Poynting flux can easily be evaluated.
  Since the compact source is of size $\sim 10^{6}$ cm,  magnetic
   fields $\sim 10^{15}$ G are needed to produce the required energy output of the GRB.

We apply  equation (1) to a rapidly rotating black hole, assumed to
be the inner engine of a popular model of the GRB \citep{pir04}. The
magnetic field in the vicinity of the black hole is obtained, using
$r \sim 10^{6}$ cm, from (3). The dimensionless spin parameter
$\alpha$ of the GRB is defined as $Jc/GM^{2}$. We then obtain the
magnetic field for a GRB in terms of the spin parameter $\alpha$
from (1):
\begin{equation}
B = \frac{G^{3/2}M^{2}\alpha\beta}{c^{2}r^{3}}\approx 225
\frac{(M/M_{\odot})^{2}}{(r/R_{\odot})^{3}} \alpha \beta \ \ G
\end{equation}
The NMGEC prediction from (7),  using  $\alpha \sim 1$, $\beta \sim$
0.1, r $\sim 10^{6}$ cm,  and M $\sim 2.5 M_{\odot}$ is B $\sim
10^{15}$ Gauss, in good  agreement with  the required field.

\section{Conclusions and Discussion}

 Observations indicate the presence of intense magnetic fields in  quasars and
  gamma-ray bursts (GRBs).
 Standard astrophysical theories have  difficulty in explaining them. We evaluated
 the magnetic fields
 predicted by non-minimal gravitational-electromagnetic coupling (NMGEC) for these objects.
 We showed that for typical
values of moments of inertia, radii, and periods for rapidly
rotating newly-born neutron stars, the NMGEC theory  predicts the
required magnetic fields.

The accretion disk magnetic field in  quasars in the region $\sim
10^{3} r_{g}$, where polarized optical radiation is generated, is
estimated to be on the order of a thousand G. For a maximally
rotating black hole in this region, NMGEC predicts this field.

In GRBs a magnetic field $\sim 10^{15}-10^{16}$ G is required to produce the
Poynting flux needed to supply the energy observed. This field is
predicted by NMGEC to exist   outside a rapidly rotating black hole
of several solar mass.

It is not easy to produce astrophysically intense magnetic
 fields. We showed here that such fields are
 predicted naturally by rapidly rotating neutron stars and black
 holes by  NMGEC. If such intense fields are definitely proven to
  exist, it would give  support for the NMGEC theory.

\section*{Acknowledgments}

R.S.S. thanks the Brazilian agency FAPESP for financial support
(04/05961-0). R.O. thanks FAPESP (06/56213-9) and the Brazilian
agency CNPq (300414/82-0) for partial support.

 \clearpage

\end{document}